\documentclass[11pt,a4paper]{article}
\usepackage{cite}
\usepackage{amsmath, amsthm, amssymb,slashed}
\usepackage{ifpdf}
\ifpdf
  \usepackage[pdftex]{graphicx}
  \usepackage{epstopdf}
\else
  \usepackage[dvips]{graphicx}
\fi
\usepackage{fancyhdr}
\usepackage{mathbbol}
\usepackage{url}
\usepackage{color}
\usepackage{bbm}
\usepackage{dsfont}
\usepackage{bm}
\usepackage{amsfonts}

\def\be{\begin{equation}}
\def\ee{\end{equation}}
\def\bea{\begin{eqnarray}}
\def\eea{\end{eqnarray}}

\parskip=\baselineskip

\def\1{{\bf 1}}
\def\2{{\bf 2}}
\def\3{{\bf 3}}
\def\4{{\bf 4}}

\font\teneurm=eurm10 \font\seveneurm=eurm7 \font\fiveeurm=eurm5
\newfam\eurmfam
\textfont\eurmfam=\teneurm \scriptfont\eurmfam=\seveneurm
\scriptscriptfont\eurmfam=\fiveeurm

\font\teneusm=eusm10 \font\seveneusm=eusm7 \font\fiveeusm=eusm5
\newfam\eusmfam
\textfont\eusmfam=\teneusm \scriptfont\eusmfam=\seveneusm
\scriptscriptfont\eusmfam=\fiveeusm

\font\tencmmib=cmmib10 \skewchar\tencmmib='177
\font\sevencmmib=cmmib7 \skewchar\sevencmmib='177
\font\fivecmmib=cmmib5 \skewchar\fivecmmib='177
\newfam\cmmibfam
\textfont\cmmibfam=\tencmmib \scriptfont\cmmibfam=\sevencmmib
\scriptscriptfont\cmmibfam=\fivecmmib

\begin{document}
\begin{titlepage}
\begin{flushright}

\end{flushright}
\vskip 1.5in
\begin{center}
{\bf\Large{Comment on `Note on $X(3872)$ production at hadron colliders\\\vspace{0.5truecm} and its molecular structure'}}
\vskip 0.5cm {A. Esposito$^a$, B. Grinstein$^b$, L. Maiani$^c$, F. Piccinini$^d$, A. Pilloni$^e$,  A.D. Polosa$^f$, V. Riquer$^c$} \vskip 0.05in 
{\small{ 
\textit{$^a$Center for Theoretical Physics and Dept. of Physics, Columbia University, New York, NY, 10027, USA}\\
\textit{$^b$Department of Physics, University of California, San Diego, La Jolla, California 92093-0315, USA}\\
\textit{$^c$CERN, Theory Department, Geneva 1211, Switzerland}\\
\textit{$^d$INFN Sezione di Pavia, Via A. Bassi 6, 27100 Pavia, Italy}\\
\textit{$^e$ Theory Center, Thomas Jefferson National Accelerator Facility, Newport News, VA 23606, USA}\\
\textit{$^f$Dipartimento di Fisica and INFN, Sapienza Universit\`a di Roma, P.le Aldo Moro 2, I-00185 Roma, Italy}\\ 
}
}
\end{center}
\vskip 0.5in
\baselineskip 16pt
\begin{abstract}
We briefly comment on the paper by Albaladejo et al.~[arXiv:1709.09101], rejecting its conclusions. 
\end{abstract}
%
\end{titlepage}

The inequality (1) of the comment by
Albaladejo {\it et al.}~\cite{Albaladejo:2017blx}  was introduced in Ref.~\cite{Bignamini:2009sk} in order to estimate the upper bound of the $X(3872)$
prompt production cross section at CDF without resorting to hadronic models for the 
calculation of the amplitude $\Psi(\bm p)=\langle X| D^0 \bar D^{*0}(\bm p)\rangle$.  We are still convinced that this is the correct way to proceed and we do not see any progress made in~\cite{Albaladejo:2017blx}.

To the effect of computing the cross section mentioned, we used the uncertainty principle to
estimate the size of the allowed ball ${\cal R}$ in the space of relative $\bm p$ momenta
in the center of mass of the generic $D^0 \bar D^{*0}$ pair produced in a $pp(\bar p)$ collision. The size of ${\cal R}$ must be compatible with the `coalescence' of a meson pair into a loosely bound meson molecule. 
Hence  the radius of the ${\cal R}$ ball is determined {\it a priori} and equation (3) in~\cite{Bignamini:2009sk}, once ${\cal R}$ is determined, simply reads
\bea
\sigma(p\bar p\to X)&\simeq& \left|\,\int_{{\cal R}}  \Psi(\bm p)\, \langle D^0\bar D^{*0}(\bm p)  |\bar p p\rangle d^3p\,\right|^2\lesssim \,
\int_{{\cal R}}  \left|\Psi(\bm p)\right|^2  d^3p
\int_{{\cal R}}  \left| \langle D^0\bar D^{*0}(\bm p)  |\bar p p\rangle\right|^2 d^3p\notag\\
&\lesssim&\int_{{\cal R}}  \left| \langle D^0\bar D^{*0}(\bm p)  |\bar p p\rangle\right|^2  d^3p.
\label{xsect}
\eea
Even if it were known how to compute from first principles the amplitude  
$\Psi(\bm p)$ (which is not the case), the size of the ${\cal R}$ ball in momentum space should be understood on the basis of physical
arguments as the ones reported below, suggesting its  radius to be $\bar p\sim20$~MeV. These arguments  involve square 
moduli of the amplitudes only. 
 
We first briefly analyze the case of the deuteron, where more experimental information is
available.   The attractive Yukawa potential can be parameterized as
\be
V=-g\,\frac{e^{-r/r_0}}{r}
\ee
where $r_0\sim 1/m_\pi=1.4$~fm and 
\be
g= \frac{f_{\pi N}^2}{4\pi}
\ee
with $f_{\pi N}\approx 2.1$, as can be computed solving the Schr\"odinger problem to get a binding energy of 2.2~MeV\footnote{The following results are also found using a  square well potential and well documented in the literature.}.  The quantum version of the virial theorem is
\be
2\overline T=\Big(\Psi, \,\sum_{i=1}^3 r_i\frac{\partial V}{\partial r_i}\Psi\Big)=-\overline V + \frac{g}{r_0}\, \overline{e^{-r/r_0}}
\ee
therefore the mean $\overline E=\overline T+\overline V$ is 
\be
\overline E=-\frac{\bar p^2}{2 \mu} + \frac{g}{r_0}\, \overline{e^{-r/r_0}}
\label{virial}
\ee
where $\mu$ is the reduced mass of the bound state.

Both the radius, $\overline r=2.1$~fm~\cite{Pohl:2016glp}, and the binding energy,
$\overline E\simeq -2.2$~MeV, are known for the deuteron.  Eq.~\eqref{virial} then gives
${\bar p}_D\approx 105$~MeV. It is this that determines the radius of the
ball ${\cal R}$,  significantly   smaller  than the value given in~\cite{Albaladejo:2017blx}, $\bar p\gtrsim 300$~MeV.

The $X(3872)$ is a more extreme case. The binding energy is found below $|\overline E|\lesssim 0.1 $~MeV
and, as commented in a large number of papers (see reviews~\cite{Esposito:2014rxa} and
references therein), for such a small binding energy the expected size of the 
state is $\bar r\gtrsim 1/\sqrt{2\mu |\overline E|} \sim 10$~fm. This allows us to neglect the second term in~\eqref{virial}, finding
${\bar p}\approx 20$~MeV, in agreement with the radius of the region ${\cal R}$ used
in~\cite{Bignamini:2009sk}. We have checked that the value of $g$ cannot be  large enough  to spoil the  approximation used. 
This can  be done either extracting $g$ from $D^*\to D \pi$ decay rate~\cite{casalbuoni}  or by solving the bound state problem.

In contrast, Albaladejo {\it et al.}~\cite{Albaladejo:2017blx}, to estimate the radius $\bar p$ of ${\cal R}$ corresponding to the production of $X(3872)$, assume that $\langle D^0\bar D^{*0}(\bm p)  |\bar p p\rangle$ in~(1) is almost 
independent of $\bm p$  
and study the quantity 
\be
I({\bar p})=
\int_{\cal R} \Psi(\bm p)\, d^3p\, 
\label{han}
\ee
seeking the $\bar p$ value such that $I({\bar p})$ gets constant from there on. As an example, the normalizable wave function for shallow bound states used by Artoisenet and Braaten~\cite{Artoisenet:2009wk} to describe the $X(3872)$ 
molecule decreases like $p^{-2}$ so that, for large $\bar p$
\be
I({\bar p}) \sim {\bar p}
\label{han2}
\ee
which indeed does not indicate any region at all. This makes the approach in~\cite{Albaladejo:2017blx}  totally useless.
To circumvent this obvious  problem, a cutoff $\Lambda$ is introduced to manipulate the wave-function $\Psi(\bm p)\to \Psi_\Lambda(\bm p)$ in~\eqref{han}. 

This cannot be the way of determining 
the size of ${\cal R}$ which, in our view, must be obtained with  arguments  involving the binding energy and the interaction coupling constant, independently of any educated  guesses on the explicit form of $\Psi(\bm p)$.  One should also notice that the {\it ad hoc} treatment of the cutoff introduces a change of sign in $\Psi(\bm p)$ depending on $\Lambda$ (see Fig.~1 of Albaladejo {\it et al.}~\cite{Albaladejo:2017blx}) --- as if the amplitude for projecting the state $|D\bar D^*\rangle$ onto the observed $|X\rangle$ could go through some spurious zeroes.  Also the $S$-wave wave function of the deuteron  in Fig.~2 of~\cite{Albaladejo:2017blx},  displays zeroes for some particular choices of the cutoff and the model, in contradiction with the fact that the ground state wave function should not present any node.  

Disregarding for a moment all these adverse considerations on~\cite{Albaladejo:2017blx}, the bare minimum one can conclude from it, is that  deuterons should be produced equally or more abundantly than molecular $X(3872)$ resonances. For a definitive test one should then compare the production of these two particles at high transverse momenta, say, $p_\perp> 15$ GeV, where the  $X(3872)$ is copiously observed at CMS. 
  
  Extrapolation of the fits shown in
  \cite{Esposito:2015fsa} suggest an extremely low deuteron production cross section at
  high transverse momenta, in agreement with our estimate of the size ${\bar p}$ of
  the deuteron as a $pn$ molecule and in contrast with the large observed cross-section for $X(3872)$ production. There are no currently available
  data for deuteron production at such high transverse momenta, however measurements might be possible at ALICE and LHCb run II.

 \emph{This paper was prepared at the request of a journal editor}.

\bibliographystyle{unsrt}

\end{document}